\documentclass[12pt]{article}
\usepackage{amsmath,amssymb,amsthm,color}
\usepackage{graphicx}
\usepackage{cite}
\usepackage{epsfig}
\usepackage{float}
\renewcommand{\baselinestretch}{1.66}

\begin{document}

\title {Coulomb excitations of monolayer germanene \\ }

\author{
\small Po-Hsin Shih$^{a}$, Yu-Huang Chiu$^b$, Jhao-Ying Wu$^c$, Feng-Lin Shyu$^d$, Ming-Fa Lin$^{a,*}$ $$\\
\small  $^a$Department of Physics, National Cheng Kung University, Tainan, Taiwan 701 \\
\small  $^b$Department of Applied Physics, National Pingtung University, Pingtung , Taiwan 900\\
\small  $^c$Center for General Education, National Kaohsiung Marine University, Kaohsiung , Taiwan 811 \\
\small  $^d$Department of Physics ROCMA, Kaohsiung, Taiwan 830 \\
 }

\renewcommand{\baselinestretch}{1.66}
\maketitle

\renewcommand{\baselinestretch}{1.66}

\begin{abstract}

The feature-rich electronic excitations of monolayer germanene lie in the significant spin-orbital coupling and the buckled structure.
The collective and single-particle excitations are diversified by the magnitude and direction of transferred momentum, the Fermi energy and the gate voltage.
There are four kinds of plasmon modes, according to the unique frequency- and momentum-dependent phase diagrams.
They behave as two-dimensional acoustic modes at long wavelength.
However, for the larger momenta, they might change into another kind of undamped plasmons, become the seriously suppressed modes in the heavy intraband e-h excitations, keep the same undamped plasmons, or decline and then vanish in the strong interband e-h excitations.
Germanene, silicene and graphene are quite different from one another in the main features of the diverse plasmon modes.\\

\end {abstract}
\renewcommand{\baselinestretch}{1.66}
\maketitle

\newpage
\vskip 0.6 truecm
\par\noindent
{\bf 1. Introduction}
\vskip 0.3 truecm

The layered IV-group condensed-matter systems have stirred a lot of experimental \cite{NM:6-770,PRL:102-808,PRB:81-406,NANO:7-620,CM:25-34,PRL:98-197403,PRL:110-076801,SOC:146-351,SCI:312-5777,JPCB:108-912,NATURE:438-201,NL:7-711,ASS:291-113,SCI:327-662,NN:10-227} and theoretical \cite{RMP:81-109,PRB:46-804,PR:71-622,PRB:75-418,PRB:82-406,NANO:19-035712,PRB:90-205417,PRL:98-806,PRL:95-801,APL:102-412,PRL:107-802,PRB:77-313,PRB:73-427} studies, mainly owing to the nano-scaled thickness and hexagonal symmetry.
They are very suitable for studying the unique two-dimensional (2D) physical phenomena, such as, the Dirac-cone band structure \cite{RMP:81-109,PRB:46-804,PR:71-622,NM:6-770,PRL:102-808}, the acoustic plasmons \cite{PRB:81-406,NANO:7-620,CM:25-34,PRB:75-418,PRB:82-406}, the quantized Landau levels \cite{PRL:98-197403,PRL:110-076801,NANO:19-035712,PRB:90-205417}, the ultrahigh carrier mobility \cite{SOC:146-351,SCI:312-5777,JPCB:108-912,PRL:98-806}, the quantum spin Hall effect \cite{NATURE:438-201,PRL:95-801,APL:102-412,PRL:107-802}, and the rich absorption spectra \cite{NL:7-711,ASS:291-113,PRB:77-313,PRB:73-427}.
Furthermore, such systems are expected to have high potential for future technological applications \cite{SCI:327-662,NN:10-227}.
Since the successful exfoliation of graphene in 2004 \cite{SCI:306-666}, germanene and silicene have been synthesized on distinct substrate surfaces, e.g., Ge on Pt(111), Au(111) \& Al(111) surfaces \cite{AM:26-820,NJP:16-2,NL:15-510} and Si on Ag(111), Ir(111) \& $\mathrm{ZrB_2}$ surfaces \cite{PRL:108-501,NL:13-685,PRL:108-5501,PSS:90-1}.
Germanene and silicene possess the buckled structures, while graphene has a planar structure.
The spin-orbital coupling (SOC) of the two formers is much stronger than of the latter.
These two characteristics will play critical roles in the essential physical properties.
This work is mainly focused on the low-frequency elementary excitations in extrinsic monolayer germanene.
A detailed comparison is also made among these three systems.

 The IV-group single-layer systems exhibit the unusual electronic properties.
The low-lying electronic structures near the K and $\mathrm{K^{\prime}}$ points (the first Brillouin zone in the inset of Fig. 1(b)) are dominated by the outermost  $p_z$ orbitals \cite{RMP:81-109,PRB:46-804}, although the buckled germanene and silicene are built from the weak mixing of  $sp^2$ and $sp^3$ chemical bondings.
They are characterized by the gap-less or separated Dirac-cone structures, depending on the existence of SOC.
Graphene, without SOC, have two linear valence and conduction bands intersecting at the Dirac point \cite{PR:71-622,NM:6-770}.
The SOC's in germanene and silicene can separate the merged Dirac points; that is, the intrinsic systems are narrow-gap semiconductors ($E_g\sim93$ meV for Ge \& $\sim7.9$ meV for Si) \cite{PRB:84-430}.
The application of gate voltage ($V_z$) further induces the splitting of spin-related energy bands \cite{PCCP:17-366,NL:12-113}.
Moreover, the band structures, with a strong wave-vector dependence, behave the anisotropic dispersions at sufficient high energy ($\sim0.2$ eV for Ge; $\sim0.3$ eV for Si) \cite{PRB:84-430}.
This is closely related to the hopping integral ($t$) of the nearest-neighbor atoms.
However, the effects of (SOC, $V_z$, $t$) on energy bands are more obvious in germanene, compared with silicene.
Such critical differences lies in the fact that the larger $4p_z$ orbitals lead to the bigger SOC and buckling, but the smaller $t$.
The main characteristics of energy bands in germanene are expected to create the diverse Coulomb excitations.

Many theoretical predictions are made on few-layer graphenes \cite{PE:54-267,RMP:81-109,JAP:106-711,NJP:8-318,SCIR:3-368,PRB:87-440,ACSN:5-026,PRB:74-406,PRB:62-508,PRB:34-979,JPSJ:69-607}, but only some studies on mono-layer silicene and germanene.
The low-frequency plasmons due to free carriers in conduction and/or valence bands can be generated by the temperature \cite{PE:54-267,JPSJ:69-607,ACSN:5-026}, interlayer atomic interactions \cite{PRB:74-406,SCIR:3-368,PRB:87-440}, doping \cite{PRB:34-979,JAP:106-711,NJP:8-318}, magnetic field \cite{ACSN:5-026}, and electric field \cite{SCIR:3-368,PRB:87-440}.
As for layered graphenes, the frequency, number and intensity of plasmon modes are very sensitive to the changes in the stacking configuration \cite{PRB:74-406,PE:54-267} and layer number \cite{PRB:62-508,PE:54-267}.
The single-layer graphene and silicene can exhibit an acoustic plasmon with a square-root dependence on the transferred momentum under the finite temperature \cite{NJP:16-2014,JPSJ:69-607} and Fermi energy \cite{PRB:34-979,PRB:89-410,NJP:8-318,JAP:106-711}.
Furthermore, the excitation spectra of the buckled silicene are enriched by $V_z$ \cite{PRB:89-410,RSC:5-912,PRB:90-142}.
In addition, the $\pi$ and $\pi\,+\sigma$ plasmons, which, respectively, come from the valence $\pi$ and $\pi\,+\sigma$ electrons are also investigated in detail \cite{PRB:62-508}.
On the experimental side, the high-resolution electron-energy-loss spectroscopy (EELS)\cite{PRB:83-403,PRB:87-447,PRB:80-410,PRB:77-406,PRL:100-803,PRB:88-433,PRB:85-440,APL:94-106} has been utilized to examine the electronic excitations in layered graphenes.
The acoustic, $\pi$ plasmons and $\pi\,+\sigma$ plasmons are identified to appear at the low ($\sim0.1-1$ eV) \cite{PRB:83-403,PRB:87-447}, middle ($\sim5-7$ eV) and high frequencies ($14-18$ eV) \cite{PRB:80-410,PRB:77-406,PRB:88-433}.
It is also noticed that the inelastic light scattering spectroscopy \cite{PRB:61-517,PRL:82-163,RMP:79-175,PRB:79-105} is successful in the verifications of the lower-frequency Coulomb excitations in doped semiconductors, e.g., $\sim$0.01-0.1 eV plasmons \cite{PRB:61-517,PRL:82-163}.

The tight-binding model, with the SOC, is used to calculate the energy bands of monolayer germanene, and the random-phase approximation (RPA) to study the $\pi$-electronic Coulomb excitations.
The dependence on the magnitude (q) and direction ($\theta$) of transferred momentum, the Fermi energy ($E_F$) and the gate voltage is explored in detail.
This work shows that there exist the diverse frequency-momentum phase diagrams, directly reflecting the significant SOC and buckled structure.
The single-particle excitations (SPEs) are greatly enriched by (q, $\theta, E_F, V_z$), so that germanene is predicted to have four kinds of collective excitation modes.
The main features of these plasmons are mainly determined by the strength of intraband and interband Landau damping and the existence of extra modes.
Germanene quite differs from silicene and graphene in excitation spectra. The theoretical predictions could be verified from the experimental measurements using the EELS \cite{PRB:83-403,PRB:80-410,PRB:77-406,PRL:100-803,PRB:88-433,PRB:85-440,APL:94-106} and the inelastic light scattering spectroscopy \cite{PRB:61-517,RMP:79-175,PRB:79-105}.

\vskip 0.6 truecm
\par\noindent
{\bf 2. The tight-binding model}
\vskip 0.3 truecm

Monolayer germanene has a low buckled structure with the weak mixing of $sp^2$ and $sp^3$ bondings; furthermore, the low-lying electronic structure is dominated by $4p_z$ orbitals.
There are two equivalent sublattices of A and B (Fig. 1(a)), being separated by a distance of $l=0.66\AA $.
Each sublattice has a hexagonal lattice with Ge-Ge bond length $b=2.32 \AA$.
Moreover, the spin-orbital coupling plays a critical role in the electronic properties.
The Hamiltonian built from the sub-space spanned by the four spin-dependent tight-binding functions is expressed as

\begin{equation}
H=-t\displaystyle\sum_{\langle i,j \rangle \alpha}c^{\dagger}_{i\alpha}c_{j\alpha}+\mathrm{i}\frac{\lambda}{3\sqrt{3}}\displaystyle\sum_{\langle \langle i,j \rangle \rangle \alpha \beta}\nu_{ij}c^{\dagger}_{i\alpha}\sigma^{z}_{\alpha\beta}c_{j\beta}- \mathrm{i}\frac{2}{3}\lambda_{R}\displaystyle\sum_{\langle \langle i,j \rangle \rangle \alpha \beta}\mu_{ij}c^{\dagger}_{i\alpha}(\vec{\sigma}\times\hat{d}_{ij})c_{j\beta}+\sum_{i \alpha}\zeta_{i}V^{i}_{z}c^{\dagger}_{i\alpha}c_{i\alpha}.
\end{equation}

\noindent The first term, in which the summation takes over all pairs $\langle i,j \rangle$ of the nearest-neighbor lattice sites, is the kinetic energy with the hopping integral of $t$ = 1.31 eV \cite{PRB:84-430}.
$c^{\dagger}_{i\alpha}$ $( c_{j\alpha})$ can create (annihilate) an electron with spin polarization $\alpha$ $(\beta)$ at site $i$ $(j)$.
The second term represents the effective SOC with the summation taking over all pairs $ \langle \langle i,j \rangle \rangle$ of the next-nearest-neighbor sites and $\lambda$ = 46.3 meV.
 $\vec {\sigma} = (\sigma_x, \sigma_y, \sigma_z) $ is the Pauli spin matrix.
$\nu_{i,j}\hat{z} = ( \vec{d_i} \times \vec{d_j} )/ |\vec{d_i} \times \vec{d_j} |$, where $\nu_{i,j}=+1$ and $-1$, respectively, correspond to the  anti-clockwise and clockwise cases from the cross product of the two nearest-neighbor bonding vectors $\vec{d_i}$ and $\vec{d_j}$.
The third term denotes the Rashba SOC with $\lambda_R = 10.7$ meV, $u_{i,j} = + 1(-1)$ for the A (B) lattice sites, and $\hat{d}_{ij}$ is the unit vector connecting two sites $i$ and $j$ in the same sublattice.
The fourth term stands for the significant sublattice potential with $\zeta_{i} = + 1(-1)$ corresponding to the A (B) sites.
The diverse relationships among the orbital bondings, the spin-orbital interactions, and the electric field can generate the unusual energy bands and thus enrich the Coulomb excitations.

\begin{figure}[H]
\centering
\includegraphics[width=0.8\linewidth]{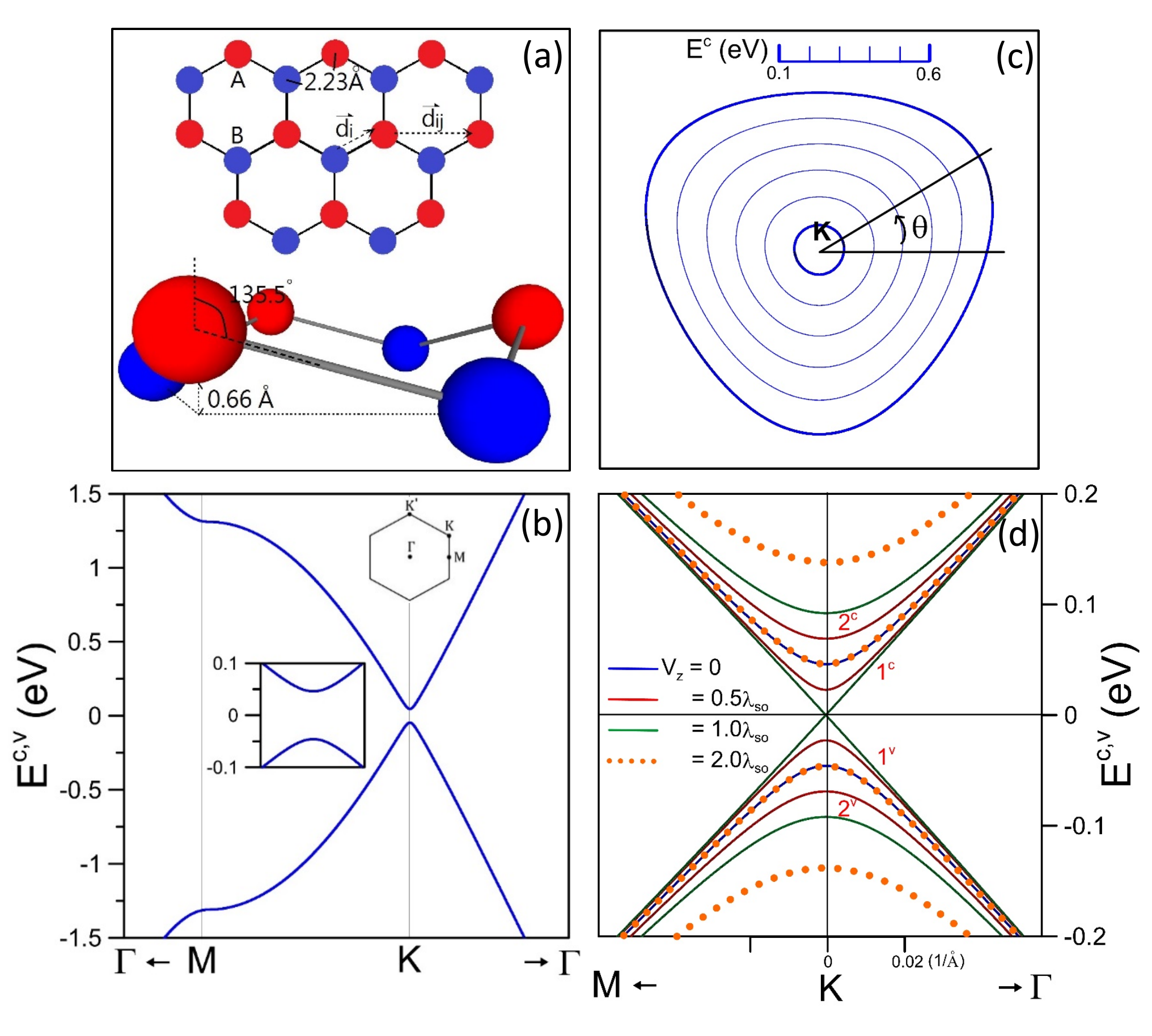}
\caption{Monolayer germanene has (a) a buckled structure, (b) a pair of valence and conduction bands along the high symmetry points, (c) the $V_z$-induced splitting energy bands near the Fermi level, and (d) the anisotropic energy contours measured from the K point.
Also shown in the insets of (b) are the separated Dirac points and the first Brillouin zone.}
\label{Figure 1}
\end{figure}

When monolayer germanene is perturbed by an external time-dependent Coulomb potential, all the valence and conduction electrons will screen this field and thus create the charge redistribution.
The effective potential between two charges is the sum of the external potential and the induced one arising from the screening charges.
The dynamical dielectric function, defined as the ratio between the bare and effective Coulomb potentials, under the RPA is given by \cite{PRB:34-979}

\begin{equation}
\begin{aligned}
\epsilon(q,\theta,\omega)=\epsilon_{0}-V_{q}\displaystyle\sum_{s,s'=\alpha,\beta}\displaystyle\sum_{h,h'=c,v}\int_{1stBZ}\frac{\mathrm{d}k_{x}\mathrm{d}k_{y}}{(2\pi)^{2}}\vert \langle s';h';\vec{k}+\vec{q} \vert \mathrm{e}^{\mathrm{i}\vec{q}\cdot\vec{r}}\vert s;h;\vec{k}\rangle\vert^{2} \\
\times \frac{f(E^{s',h'}(\vec{k}+\vec{q}))-f(E^{s,h}(\vec{k}))}{E^{s',h'}(\vec{k}+\vec{q})-E^{s,h}(\vec{k})-(\omega+\mathrm{i}\Gamma)}.
\end{aligned}
\end{equation}

\noindent $q_x=qcos\theta$ and $q_y=qsin\theta$, where $\theta$ is the angle between $\vec{q}$ and $\hat{k_x}$ (along $\Gamma$M in the inset of Fig. 1(b)).
The $\pi$-electronic excitations are described by the magnitude and direction of the transferred momentum and excitation frequency.
$0^{\circ}\le\theta\le30^{\circ}$ is sufficient to characterize the direction-dependent excitation spectra because of the hexagonal symmetry.
$\epsilon_0=2.4$ is the background dielectric constant, and $V_{q}=2\pi\mathrm{e}/q$ is the bare Coulomb potential energy.
 $f(E)=1/\{1+\mathrm{exp}[(E-\mu)k_BT]\}$ is the Fermi-Dirac distribution, where $k_B$ is the Boltzmann constant, and $\mu$ the chemical potential corresponding to the highest occupied state energy in the metallic systems (or the middle energy of band gap in the semiconducting systems).
$\Gamma$ is the energy width due to various de-excitation mechanisms, and is treated as a free parameter in the calculations.

\vskip 0.6 truecm
\par\noindent
{\bf 3. Results and Discussion}
\vskip 0.3 truecm

Germanene exhibits the feature-rich band structure which strongly depends on the spin-orbital coupling, the wave vector, and the gate voltage.
The unoccupied conduction band is symmetric to the occupied valence one about the zero energy.
Electronic states in the presence of SOC are doubly degenerate for the spin freedom (Fig. 1(b)).
They have up- and down-dominated spin configurations simultaneously.
Two distinct spin configurations can make the same contribution to the Coulomb excitations.
Moreover, the SOC can generate the separation of Dirac points and thus induce an energy spacing of $E_D=93$ meV near the K point (the blue curves).
The energy bands have parabolic dispersions near the band-edge states and then gradually become linear ones in the increase of wave vector.
However, at higher energy $(|E^{c,v}|> 1)$, they recover into parabolic ones and have a saddle point at the M point with very high density of states .
It is noticed that only the low-lying states with $|E^{c,v}|<0.2$ eV exhibit the isotropic energy spectra, as indicated from constant-energy contours in Fig. 1(c).

The main features of low-lying energy bands are easily tuned by a gate voltage (or a perpendicular electric field).
The Coulomb potential energy difference between the A and B sublattices further results in the destruction of mirror symmetry about the $z=0$ plane.
The spin-dependent electronic states are split; that is, one pair of energy bands changes into two pairs of ones.
These conduction and valence bands are, respectively, denoted by $1^{c,v}$ and $2^{c,v}$ in Fig. 1(d), e.g., $E^{c,v}$ at $V_z=0.5 \lambda_{so}$ (the red curves), where the effective SOC is $\lambda_{so} =E_D/l=141$ meV $/\textup{\AA}$.
As for the electronic states near the K point, the pair of energy bands for the spin-down-dominated configuration is relatively close to zero energy.
However, the opposite is true for those near the $\mathrm{K^{\prime}}$ point.
As $V_z$ grows, the first pair gradually approaches to zero energy and $E_D$ is getting smaller.
$E_D$ is vanishing at $V_z=\lambda_{so}$ (the green curves), where the intersecting linear bands are revealed, or the electronic structure has a pair of zero-spacing linear bands.
With the further increase of $V_z$ (the dashed orange curves), the energy spacing of parabolic valence and conduction bands will be opened and enlarged.
On the other hand, the second pair of energy bands for the spin-up-dominated configuration is away from zero energy.
Apparently, two splitting spin-dependent configurations are expected to greatly diversify the single-particle and collective excitations.

\begin{figure}[H]
\centering
\includegraphics[width=0.8\linewidth]{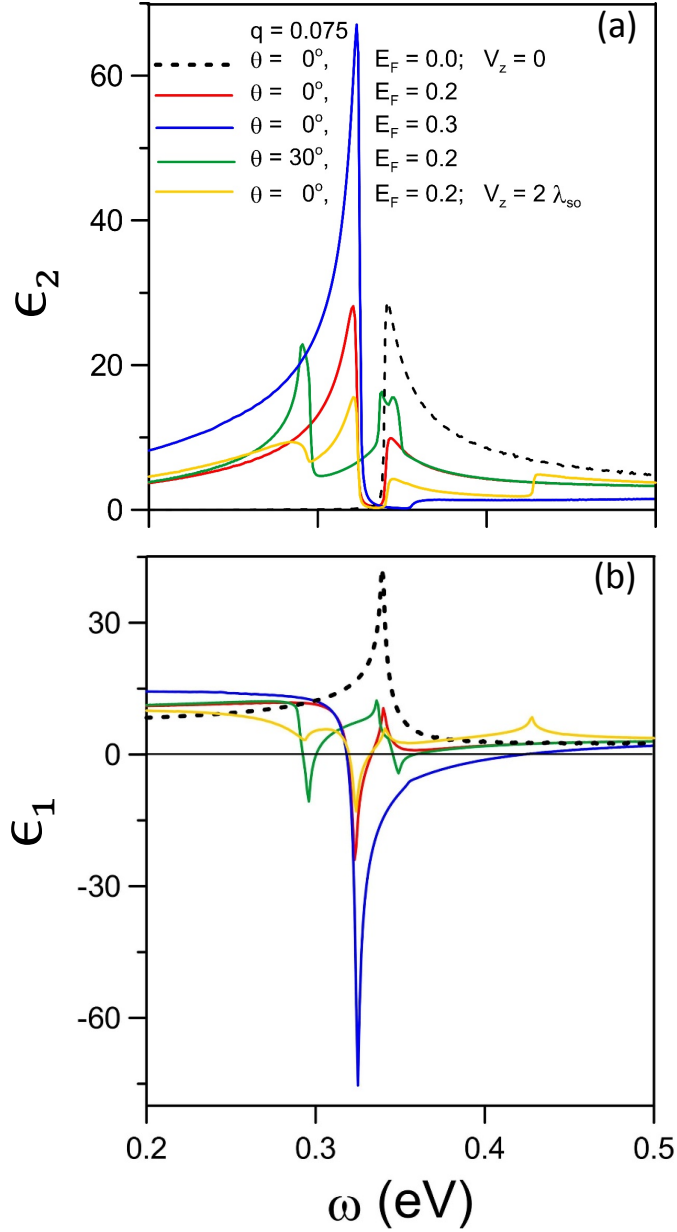}
\caption{(a) The imaginary part and (b) the real part of the dielectric function at $q=0.075 \textup{\AA}^{-1}$ for the distinct $E_F$'s, $\theta$'s and $V_z$'s'.}
\label{Figure 2}
\end{figure}

The SPEs are characterized by the imaginary part of the dielectric function ($\epsilon_2$).
They strongly depend on the direction and magnitude of transferred momentum, Fermi energy, and gate voltage.
The non-vanishing $\epsilon_2$ corresponds to the interband and intraband SPEs consistent with the Fermi-Dirac distribution, and the conservation of energy and momentum.
$\epsilon_2$ exhibits an asymmetric square-root divergent peak at the threshold energy $\omega_{th}^{inter} = 2\sqrt{(\lambda_{so})^2 + v_F^2 q^2 / 4}$ at $E_F=0$ and $\theta=0^{\circ}$ (the black dashed curve in Fig. 2(a); $v_F=3tbq/2$ is the Fermi velocity) \cite{PRB:34-979,PRB:84-430}.
This structure comes from the interband excitations which are associated with the valence or the conduction band-edge states (the separated Dirac points).
From the Kramers-Kronig relations between the real and the imaginary parts of the dielectric function, $\epsilon_1$ exhibits a similar divergent peak at the left-hand side (Fig. 2(b)).
It is always positive at the low-frequency range, implying that collective excitations hardly survive.

The competition between the interband and intraband excitations is created by the changes of the free carrier density.
The former are gradually suppressed by the latter in the increase of the Fermi energy.
The former and the latter could coexist at $E_F= 0.2$ eV; they, respectively, have a shoulder structure and a prominent square-root divergent peak in $\epsilon_2$ (red curve).
These two structures, which are closely related to the electronic excitations of the Fermi-momentum states, are separated by an excitation gap due to the energy spacing of Dirac points.
The strong asymmetric peak appears at $\omega_{ex}^{intra}\approx v_Fq$, directly reflecting the linear energy dispersion.
This specific excitation energy is the upper boundary of the intraband SPEs (Fig. 4(a)).
Moreover, $\epsilon_1$ exhibits the logarithmic divergent and square-root peaks, as shown in Fig. 2(b) \cite{SCIR:3-368}.
The two zero points in $\epsilon_1$, if at where $\epsilon_2$ is sufficiently small, are associated with collective excitations.
The intraband excitations become the dominating channel in the low-frequency SPE spectrum at $E_F= 0.3$ eV (blue curve).
$\epsilon_1$ and $\epsilon_2$ have the peak structures, in which the second zero point of the former is located at where the latter is rather small.
The intensity of intraband excitation peaks is getting strong with the increasing $E_F$.
However, the zero point of $\epsilon_1$ is revealed at the higher frequency, accompanied with the reduced $\epsilon_2$.
That is, the further increment of $E_F$ will reduce the e-h Landau dampings and enhance the frequency of collective excitations.

The changes in the direction of transferred momentum lead to the anisotropic SPEs.
The constant-energy contours are vertically flipped between the K (Fig. 1(c)) and $\mathrm{K^{\prime}}$ points.
The energy variation measured from them along the direction of $\theta=30^{\circ}$ is, respectively, smaller and bigger, compared with the  $\theta=0^{\circ}$ case.
In the vicinity of the K ($\mathrm{K^{\prime}}$) point, the highest intraband and the lowest interband excitation energies become lower (higher).
As a result, each above-mentioned structure is changed into double structures at $E_F=0.2$ eV and $q=0.075 \textup{\AA}^{-1}$ (red curve versus green curve in Fig. 2(a)).
Furthermore, the higher-frequency intraband peak might overlap with the first interband shoulder, instead of the excitation gap associated with the K point.
The distinct excitations near the two valleys are also reflected in $\epsilon_1$, e.g., the change of special structures.
With the increasing $\theta$, the widened frequency range of the e-h damping is expected to significantly affect the plasmon modes.

The gate voltage can split energy bands and thus diversify the channels of SPEs.
There are eight kinds of excitation channels, in which two and six kinds, respectively, belong to the intraband and the interband excitations.
Specially, the four interband excitation channels, $2^v\rightarrow1^c$, $2^c\rightarrow1^c$, $1^v\rightarrow2^c$ and $1^c\rightarrow2^c$ in the low-frequency excitation range are negligible, mainly owing to the almost vanishing Coulomb matrix elements (the square term in Eq. (2)).
The $V_z$-dependent dominating e-h excitations come from the intraband excitations ($2^c\rightarrow2^c$, $1^c\rightarrow1^c$) and the interband excitations ($1^v\rightarrow1^c$, $2^v\rightarrow2^c$).
They, respectively, cause $\epsilon_2$ to exhibit two square-root divergent peaks and two shoulder structures, as shown at $V_z = 2\lambda_{so}$, $E_F=0.2$ eV, $\theta=0^{\circ}$ and $q=0.075 \textup{\AA}^{-1}$ (yellow curve).
The second peak due to the $1^c\rightarrow1^c$ channel has the excitation frequency almost independent of $V_z$.
Moreover, the threshold excitation energies of the $1^v\rightarrow1^c$ and $2^v\rightarrow2^c$ interband excitations, which mainly come from the valence Dirac point and the Fermi-momentum state, are given by $\omega_{th}^{inter}=2\sqrt{(\lambda_{so} \mp V_z)^2 + V_F^2 q^2 / 4}$ \cite{PRB:84-430}.
They will determine the enlarged boundaries of the interband SPEs (Fig. 5).
The frequency of the former shoulder structure becomes lower as $V_z$ reduces from $2\lambda_{so}$ to $ \lambda_{so}$ and then higher in the further decrease from $ \lambda_{so}$ to 0.
Differently, that of the latter is monotonically declining with the decrease of $V_z$.
However, the zero-point frequency in $\epsilon_1$, corresponding to the $1^c\rightarrow1^c$ e-h peak, hardly depends on $V_z$.
This will induce the suppression of the interband Landau dampings on plasmon modes.

\begin{figure}[H]
\centering
\includegraphics[width=0.8\linewidth]{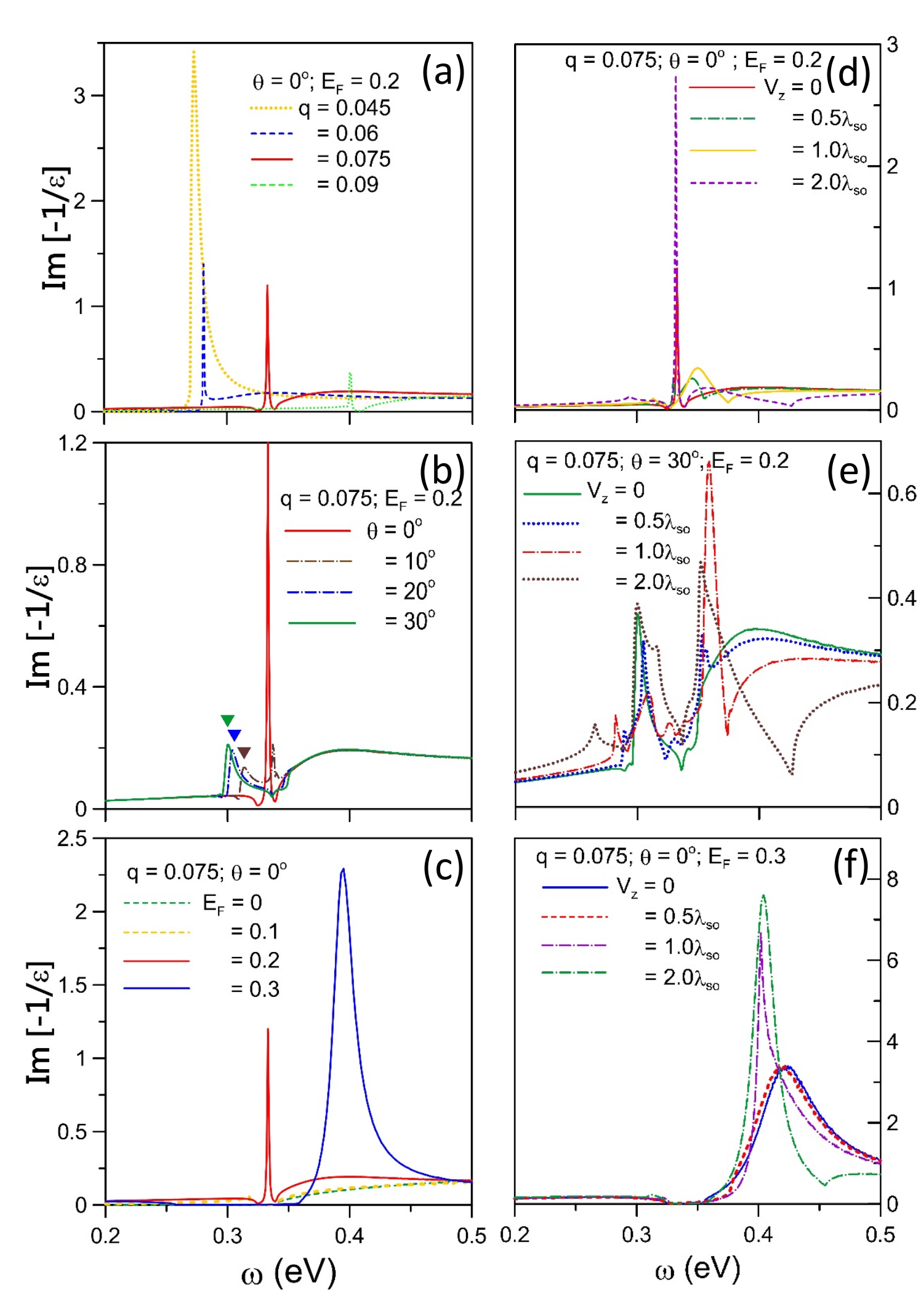}
\caption{The energy loss spectra (a) for ($\theta=0^{\circ}$, $E_F=0.2$ eV) at different $q$'s, (b) for ($q=0.075 \textup{\AA}^{-1}$, $E_F=0.2$ eV) at different $\theta$'s, and (c) for ($q=0.075 \textup{\AA}^{-1}$, $\theta=0^{\circ}$) at different $E_F$'s.
The $V_z$-dependent spectra are shown for (d) ($q=0.075 \textup{\AA}^{-1}$, $\theta=0^{\circ}$, $E_F=0.2$ eV), (e) ($q=0.075 \textup{\AA}^{-1}$, $\theta=30^{\circ}$, $E_F=0.2$ eV), and (f) ($q=0.075 \textup{\AA}^{-1}$, $\theta=0^{\circ}$, $E_F=0.3$ eV).}
\label{Figure 3}
\end{figure}

The loss function, defined as $Im[-\frac{1}{\epsilon}]$, measures the screened excitation spectrum and is useful to characterize the collective excitations.
It strongly depends on the magnitude and direction of transferred momentum.
There exists a prominent peak at $q=0.045 \textup{\AA}^{-1}$, $\theta=0^{\circ}$ and $E_F=0.2$ eV, as indicated by a yellow dashed curve in Fig. 3(a).
It corresponds to the collective excitations of free carriers in conduction bands.
With the increase of momentum, the plasmon frequency ($\omega_p$) grows because of the higher-energy zero point in $\epsilon_1$, while its intensity is reduced by the stronger interband e-h damping.
Moreover, a prominent peak is changed into a composite structure of a narrow sharp peak and a shoulder structure.
The loss spectrum is enriched by the distinct directions of momentum transfer.
An extra peak, marked by the triangle in Fig. 3(b) ($q=0.075 \textup{\AA}^{-1}$ and $E_F=0.2$), comes to exist with the increase of $\theta$.
This arises from the free carriers in the K valley and it is damped by the SPEs due to the $\mathrm{K^{\prime}}$ valley.
The enlarged frequency range of the e-h dampings will cover the excitation gap at larger $\theta$'s.
As a result, the sharp peak is seriously suppressed and only the shoulder structure due to the $\mathrm{K^{\prime}}$ valley can survive.

The variation in free carrier density could modify the SPE channels and thus drastically change the plasmon peaks in the loss spectrum.
Without free carriers, only a shoulder structure corresponding to the interband SPEs is revealed in the excitation spectrum at $q=0.075 \textup{\AA}^{-1}$ and $\theta=0^{\circ}$ in Fig. 3(c) (green curve); that is, the plasmon peak is absent.
When the free carrier density gradually grows, the loss spectrum exhibits a narrow sharp plasmon peak and a shoulder structure, being attributed to the strong interband e-h dampings.
With the further increase of $E_F$, those two structures are enhanced, and then only a prominent plasmon peak comes to exist.
This is due to the fact that the interband excitations are progressively replaced by the intraband excitations.\\

The gate voltage can diversify the spin-dependent SPE channels and thus enrich the loss spectra (Figs. 3(d)-3(f)).
The sharp plasmon peak might change into the lower and broader one under the $V_z$-enlarged SPE range (Fig. 3(d)).
This indicates the transformation between two different kinds of plasmon modes (Fig. 6(a)).
There exist more low peak structures induced by the extra excitation channels, especially for the large-$\theta$ case (Fig. 3(e)).
With the increase of $V_z$, the intensity of the sharp plasmon peak is greatly enhanced for the large $E_F$ (Fig. 3(f)).
However, the monotonous $V_z$-dependence is absent in the other cases.
The variation of plasmon intensity is determined by whether the inter-band excitation energy is away from or close to the plasmon frequency.

\begin{figure}[H]
\centering
\includegraphics[width=0.8\linewidth]{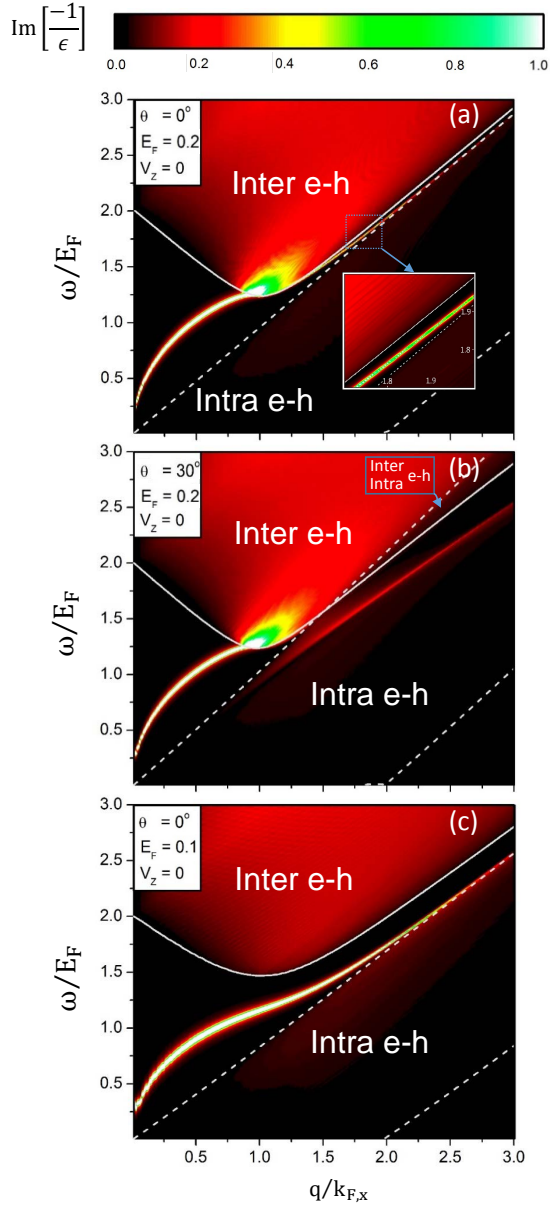}
\caption{The (momentum, frequency)-phase diagrams at (a) ($\theta=0^{\circ}$, $E_F=0.2$ eV, $V_z=0$), (b) ($\theta=30^{\circ}$, $E_F=0.2$ eV, $V_z=0$), and (c) ($\theta=0^{\circ}$, $E_F=0.1$ eV, $V_z=0$).
The intraband and interband SPE regions are, respectively, enclosed by the dashed and solid curves.
The plasmon frequency dispersions are indicated by the bright curves.
$k_{F,x}$ is the Fermi momentum along the $\Gamma$M direction.}
\label{Figure 4}
\end{figure}

Free carriers could induce the unique SPEs and plasmon modes, as clearly indicated in the $(w,q)$-dependent phase diagrams.
The boundaries of intraband and interband SPEs are mainly determined by the Fermi-momentum states and the Dirac points, as shown in Figs. 4(a)-4(c).
There exist three kinds of collective excitations with the relatively strong intensities, according to the Landau damping.
First, the free-carrier induced plasmon is undamped at small $q$ for $E_F=0.2$ eV and $\theta=0^{\circ}$.
The excitation gap, which is formed between the intraband and the interband SPEs, is responsible for this behavior (the white dashed and solid curves).
In addition, the $(q,\omega)$-range of the excitation gap grows with the increment of Dirac-point spacing.
With the increasing q, this plasmon will enter the interband excitation region and experience the gradually enhanced damping.
Specially, the partially undamped plasmon exhibits a sharp peak beyond a sufficiently large $q$; that is, this mode could survive in the frequency gap between the interband and the intraband excitation boundaries.
Secondly, at large $\theta=30^{\circ}$ (Fig. 4(b)), the larger-$q$ undamped plasmon is replaced by the seriously damped plasmon in the intraband excitation region (the red curve).
The weak plasmon intensity is due to the replacement of excitation gap by the extended and overlapped excitation regions.
Moreover, for a sufficiently low Fermi energy ($E_F=0.1$ eV in Fig. 4(c)), the plasmon remains as an undamped mode even at large $q$.
This plasmon appears in the enlarged excitation gap determined by the comparable $E_F$ and $E_D$.
The above-mentioned plasmon frequencies have the $\sqrt{q}$-dependence at long wavelength limit, a feature of  acoustic plasmon modes.
The similar mode had been observed in a 2D electron gas \cite{RMP:54-437}.

\begin{figure}[H]
\centering
\includegraphics[width=0.8\linewidth]{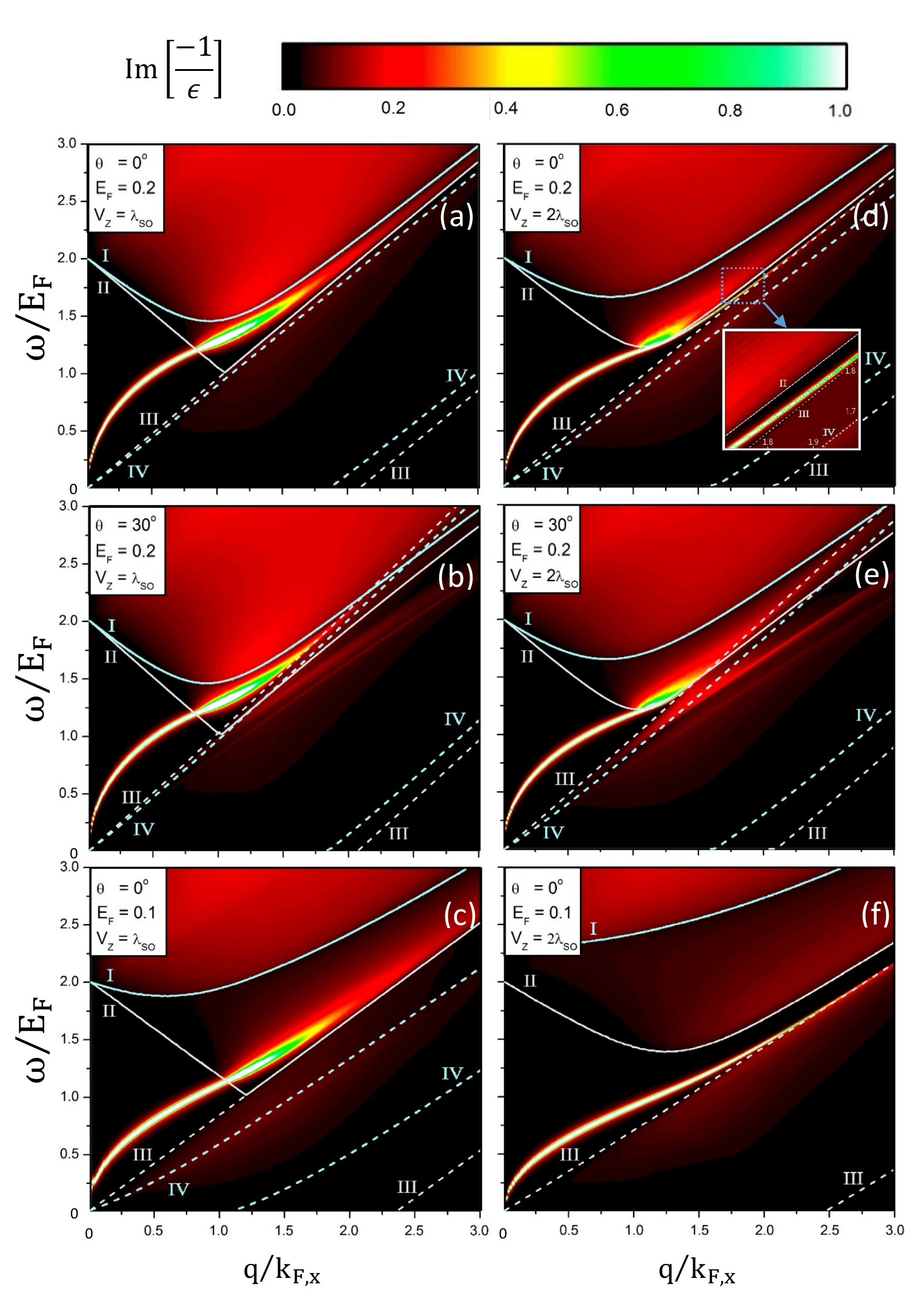}
\caption{Same plot as Fig. 4, but shown at $V_z=\lambda_{so}$ in (a), (b) \& (c) and at $V_z=2\lambda_{so}$ in (d), (e) \& (f).
I, II, III and IV, respectively, correspond to the $2^v\rightarrow2^c$, $1^v\rightarrow1^c$, $1^c\rightarrow1^c$ and $2^c\rightarrow2^c$ SPE channels.}
\label{Figure 5}
\end{figure}

The gate voltage leads to more complicated SPE boundaries because of the splitting energy bands, as revealed in Figs. 5(a)-5(f).
I, II, III and IV, respectively, represent the $2^v\rightarrow2^c$, $1^v\rightarrow1^c$, $1^c\rightarrow1^c$ and $2^c\rightarrow2^c$ SPEs.
Specially, $V_z$ can create the fourth kind of plasmon mode during the dramatic variation of energy gap.
For $V_z=\lambda_{so}$, $E_F=0.2$ eV and $\theta=0^{\circ}$, an undamped plasmon occurs at small q (Fig. 5(a)).
This plasmon is dominated by the first pair of the $V_z$-induced energy bands without energy spacing.
At larger $q$, it will compete with the interband SPEs and gradually disappears.
The damping behavior is closely related to the vanishing excitation gap and the enlarged SPE regions.
Moreover, the main characteristics of the $(w,q)$-phase diagrams are insensitive to the changes in $E_F$ and $\theta$, as indicated in Figs. 5(b)-5(c).
A similar plasmon mode is also observed in an extrinsic graphene\cite{JAP:106-711,NJP:8-318,PRB:83-403,PRB:87-447}, while germanene has the stronger intensity because of the weakened e-h damping.
With the increase of $V_z$ (Figs. 5(d)-5(f)), the main plasmon modes are similar to those of the zero-voltage ones (Figs. 4(a)-4(c)).
However, the SPE regions become more complicated, so that the extra plasmon modes could survive at phase diagrams, e.g., two weak plasmon modes heavily damped by the intraband SPEs in Fig. 5(e).

\begin{figure}[H]
\centering
\includegraphics[width=0.8\linewidth]{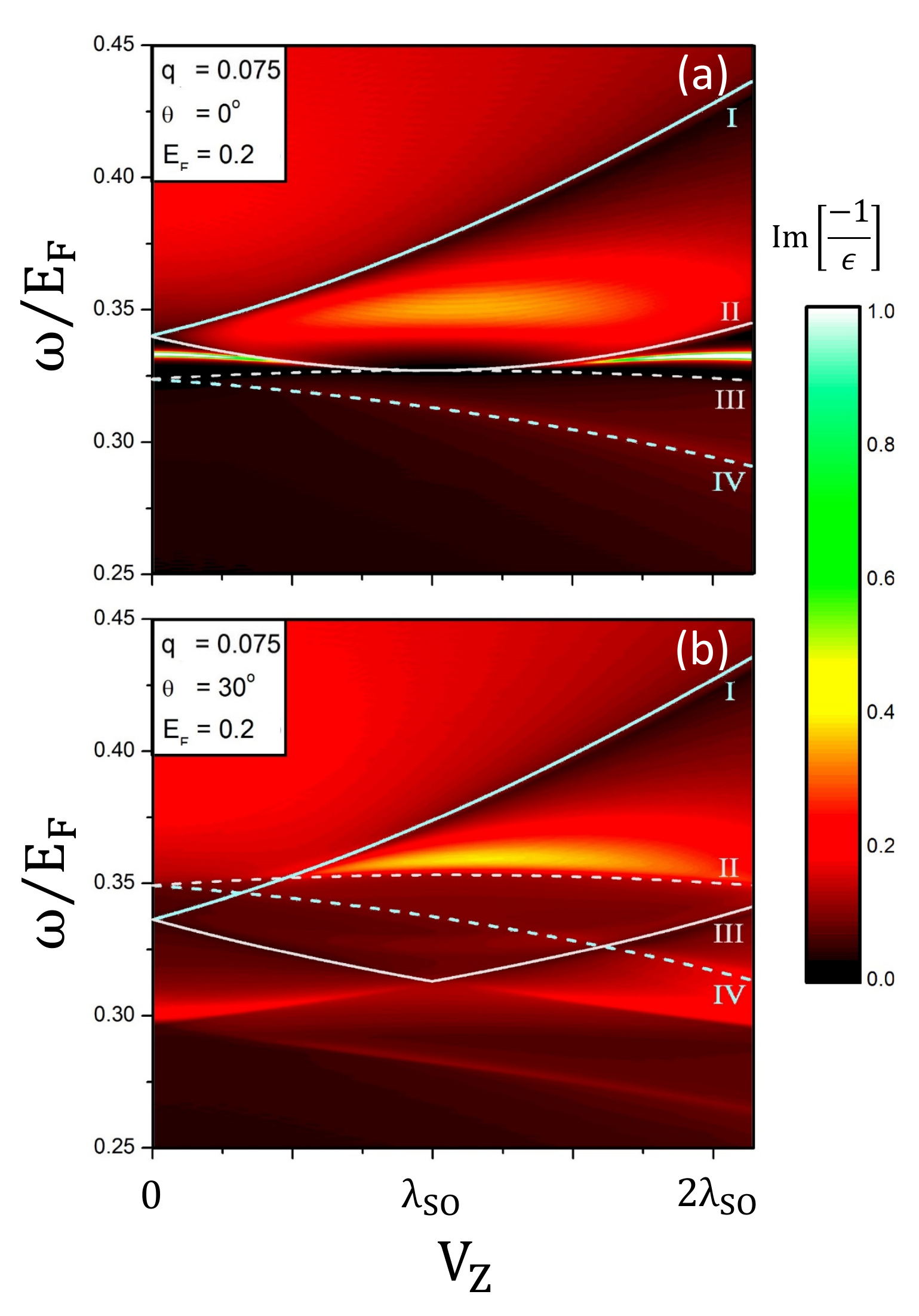}
\caption{The (voltage, frequency)-dependent phase diagrams for ($q=0.075 \textup{\AA}^{-1}$, $E_F=0.2$ eV) at (a) $\theta=0^{\circ}$, and (b) $\theta=30^{\circ}$.}
\label{Figure 6}
\end{figure}

The $V_z$-dependent excitation spectrum can provide more information on the collective and single-particle excitations.
The plasmon modes and the e-h boundaries are dramatically altered during the variation of $V_z$.
The undamped plasmon at larger $q$ will vanish within a certain range of $0.5\lambda_{so} \le V_z \le 1.5\lambda_{so}$ (Fig. 6(a)), being attributed to the serious suppression of $1^v\rightarrow1^c$ SPEs.
It is replaced by the fourth kind of plasmon which occurs in the absence of $1^c\rightarrow1^c$ SPEs.
With the further increase of $V_z$, the undamped plasmon will be recovered.
The dependence of plasmon mode on $V_z$ is very sensitive to the change in the direction of transferred momentum (Fig. 6(b)).
The undamped plasmon is absent at low and high voltages.
However, the fourth kind of plasmon can survive in the middle $V_z$-range.

\begin{figure}[H]
\centering
\includegraphics[width=0.8\linewidth]{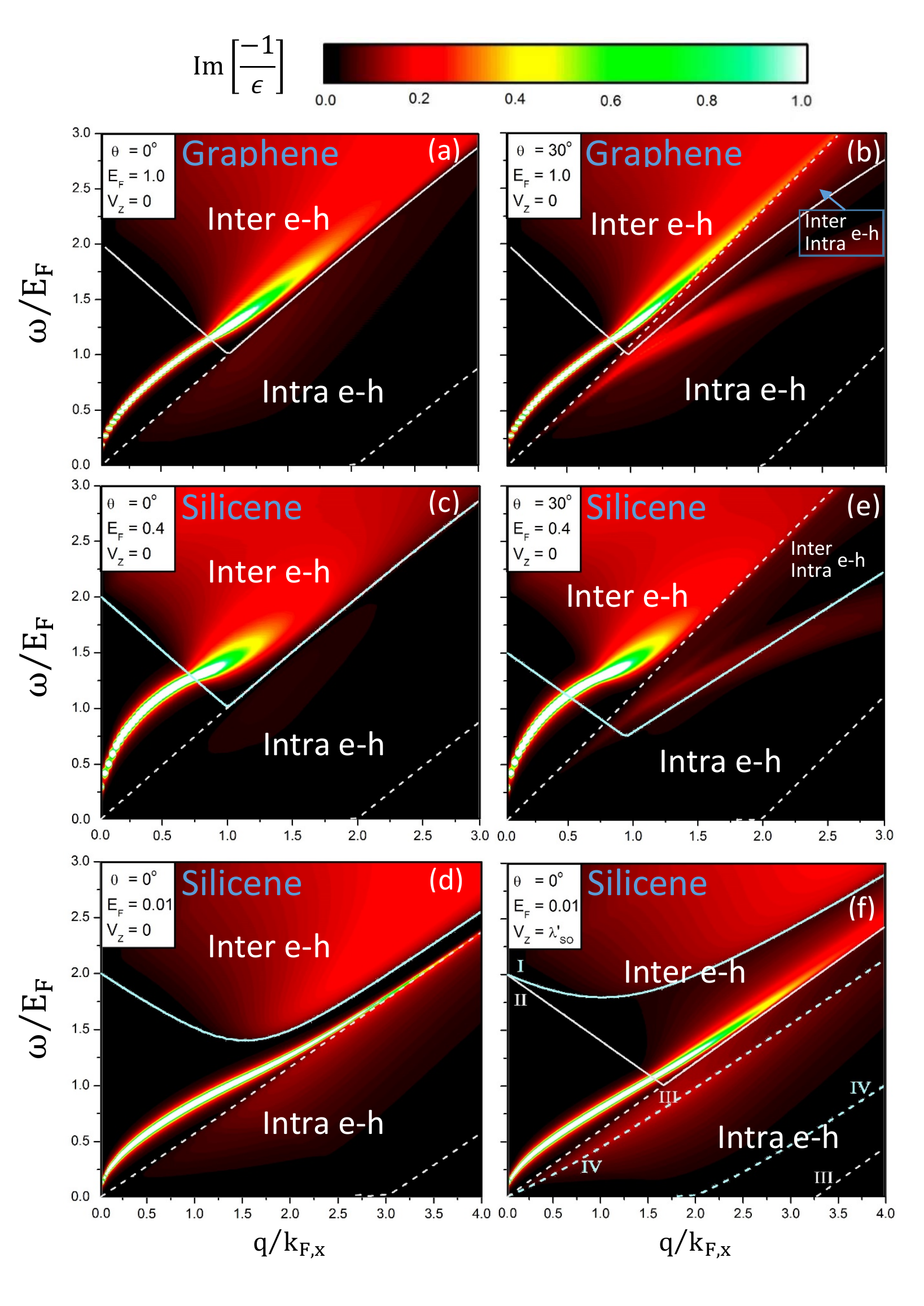}
\caption{The (momentum, frequency)-phase diagrams for graphene at (a) ($\theta=0^{\circ}$, $E_F=1.0$ eV, $V_z=0$); (b) ($\theta=30^{\circ}$, $E_F=1.0$ eV, $V_z=0$), and for silicene at (c) ($\theta=0^{\circ}$, $E_F=0.4$ eV, $V_z=0$), (d) ($\theta=30^{\circ}$, $E_F=0.4$ eV, $V_z=0$), (e) ($\theta=0^{\circ}$, $E_F=0.01$ eV, $V_z=0$); (f) ($\theta=0^{\circ}$, $E_F=0.01$ eV, $V_z=\lambda_{so}=17.5$ meV $/\textup{\AA}$).}
\label{Figure 7}
\end{figure}

There are certain important differences among germanene, silicene and graphene, mainly owing to the strength of spin-orbit interaction and the buckled structure.
In addition, the Hamiltonians of silicene and graphene are, respectively, given in Refs. 37 and 48.
For the extrinsic systems, all of them can exhibit the similar 2D plasmon modes at long wavelength limit (Figs. 7 and 5), in which the effects due to the SOI-induced energy spacing are fully suppressed.
But at larger transferred momenta, the excitation gap can create an undamped plasmon mode in germanene.
The first kind of plasmon is absent in silicene and graphene because of the rather small $E_D$'s ($<8$ meV), as shown in Figs. 7(a), 7(c); 4(a).
To observe the $\theta$-dependent plasmon modes, the required Fermi energies are, respectively, about $1.0$ eV, $0.4$ eV and $0.2$ eV for graphene, silicene and germanene (Figs. 7(b), 7(d); 4(b)).
The third kind of plasmon is revealed in germanene and silicene, while the Fermi energy needs to be very low ($\sim0.01$ eV in Fig. 7(e)) for the latter.
Also, silicene can exhibit the fourth kind of plasmon at the lower gate voltage, e.g., $V_z$ equal to the effective SOC ($\lambda_{so}=17.5 $meV$/\textup{\AA}$ in Fig. 7(f)).
In short, germanene possesses four kinds of plasmon modes only under the small variations in the Fermi energy.

The high-resolution EELS can serve as the most powerful experimental technique to explore the Coulomb excitations in 2D materials.
The experimental measurements on single- and few-layer graphenes are utilized to comprehend the collective excitations due to the free carriers, the $\pi$ electrons, and the $\pi\,+\sigma$ electrons.
They have identified the low-frequency acoustic plasmon ($\omega_p\lesssim 1$\ eV), accompanied with the interband Landau damping at larger momentum \cite{PRB:83-403}.
The interband $\pi$ and $\pi\,+\sigma$ plasmons are, respectively, observed at $\omega_p\gtrsim 4.8$ eV and $14.5$ eV; furthermore, their frequencies are enhanced by the increasing layer number \cite{PRB:80-410}.
The feature-rich electronic excitations of monolayer germanene, the four kinds of plasmon modes, the $E_F$- and $E_D$-created excitation gaps, the interband and intraband SPEs, and the $V_z$-enriched excitation spectra, are worthy of further examinations by using EELS.
The detailed measurements on the loss spectra can provide the full information on the diverse $(q, \omega)$-phase diagrams, especially for the strong $(q, \theta, E_F, V_z)$-dependence.
Also, they are useful in distinguishing the critical differences of excitation spectra among germanene, silicene and graphene.
%It should be noticed that the anisotropy distinction will be reduced to minimum at $V_z=\lambda_{so}$.This is caused by the fact that the gate voltage enlarges the SPE regions, similar to the case of the anisotropy one.

\vskip 0.6 truecm
\par\noindent
{\bf 4. Conclusion}
\vskip 0.3 truecm
In this work, we have studied the low-frequency elementary excitations of monolayer germanene within the RPA.
The calculated results show that the SPE regions are mainly determined by the Fermi-Dirac distribution, and the conservation of energy and momentum, and they are greatly enlarged by $V_z$.
Four kinds of plasmon modes are predicted to reveal in the $(q, \theta, E_F, V_z)$-dependent loss spectra; furthermore, their main features are characterized by the undamped and damping behaviors.
The differences and similarities among germanene, silicene and graphene lie in the existence of plasmon modes and excitation gaps.
Similar studies could be further extended to the middle-frequency $\pi$-electronic excitations, the few-layer germanene, and the other IV-group systems (e.g., the single-layer Sn and Pb).

 The excitation properties directly reflect the characteristics of the low-lying bands, the strong wave-vector dependence, the anisotropic behavior, the SOC-created separation of Dirac points, and the $V_z$-induced destruction of spin-configuration degeneracy.
There exists a forbidden excitation region between the intraband and interband SPE boundaries, being attributed to the Fermi-momentum and band-edge states.
The undamped plasmons could survive within this region with a prominent peak intensity.
All the plasmons due to the free conduction electrons belong to 2D acoustic modes at small q's, as observed in an electron gas.
With the increasing q, they might experience the interband damping and become another kind of undamped plasmons, change into the seriously suppressed modes in the heavy intraband damping, remain the same undamped plasmons, or gradually vanish during the enhanced interband damping.
Specifically, the first kind of plasmon modes only appears in germanene with the stronger SOC.
The fourth kind of plasmon modes in monolayer germanene are purely generated by $V_z$, while they are frequently revealed in few-layer extrinsic graphenes without external fields \cite{NJP:8-318,JAP:106-711,PRB:87-447,PRB:83-403}.
The detailed measurements using EELS could examine the diverse $(q, \omega)$-phase diagrams and verify the differences or similarities among the 2D group-IV systems.

\newpage
\vskip 0.6 truecm
\par\noindent
\begin{center}{\bf Figure Captions} \end{center}
\vskip 0.3 truecm

Fig. 1. Monolayer germanene has (a) a buckled structure, (b) a pair of valence and conduction bands along the high symmetry points, (c) the $V_z$-induced splitting energy bands near the Fermi level, and (d) the anisotropic energy contours measured from the K point.
Also shown in the insets of (b) are the separated Dirac points and the first Brillouin zone.\\

Fig. 2. (a) The imaginary part and (b) the real part of the dielectric function at $q=0.075 \textup{\AA}^{-1}$ for the distinct $E_F$'s, $\theta$'s and $V_z$'s'.\\

Fig. 3. The energy loss spectra (a) for ($\theta=0^{\circ}$, $E_F=0.2$ eV) at different $q$'s, (b) for ($q=0.075 \textup{\AA}^{-1}$, $E_F=0.2$ eV) at different $\theta$'s, and (c) for ($q=0.075 \textup{\AA}^{-1}$, $\theta=0^{\circ}$) at different $E_F$'s.
The $V_z$-dependent spectra are shown for (d) ($q=0.075 \textup{\AA}^{-1}$, $\theta=0^{\circ}$, $E_F=0.2$ eV), (e) ($q=0.075 \textup{\AA}^{-1}$, $\theta=30^{\circ}$, $E_F=0.2$ eV), and (f) ($q=0.075 \textup{\AA}^{-1}$, $\theta=0^{\circ}$, $E_F=0.3$ eV).\\

Fig. 4. The (momentum, frequency)-phase diagrams at (a) ($\theta=0^{\circ}$, $E_F=0.2$ eV, $V_z=0$), (b) ($\theta=30^{\circ}$, $E_F=0.2$ eV, $V_z=0$), and (c) ($\theta=0^{\circ}$, $E_F=0.1$ eV, $V_z=0$).
The intraband and interband SPE regions are, respectively, enclosed by the dashed and solid curves.
The plasmon frequency dispersions are indicated by the bright curves.
$k_{F,x}$ is the Fermi momentum along the $\Gamma$M direction.\\

Fig. 5. Same plot as Fig. 4, but shown at $V_z=\lambda_{so}$ in (a), (b) \& (c) and at $V_z=2\lambda_{so}$ in (d), (e) \& (f).
I, II, III and IV, respectively, correspond to the $2^v\rightarrow2^c$, $1^v\rightarrow1^c$, $1^c\rightarrow1^c$ and $2^c\rightarrow2^c$ SPE channels.\\

Fig. 6. The (voltage, frequency)-dependent phase diagrams for ($q=0.075 \textup{\AA}^{-1}$, $E_F=0.2$ eV) at (a) $\theta=0^{\circ}$, and (b) $\theta=30^{\circ}$.\\

Fig. 7. The (momentum, frequency)-phase diagrams for graphene at (a) ($\theta=0^{\circ}$, $E_F=1.0$ eV, $V_z=0$); (b) ($\theta=30^{\circ}$, $E_F=1.0$ eV, $V_z=0$), and for silicene at (c) ($\theta=0^{\circ}$, $E_F=0.4$ eV, $V_z=0$), (d) ($\theta=30^{\circ}$, $E_F=0.4$ eV, $V_z=0$), (e) ($\theta=0^{\circ}$, $E_F=0.01$ eV, $V_z=0$); (f) ($\theta=0^{\circ}$, $E_F=0.01$ eV, $V_z=\lambda_{so}=17.5$ meV $/\textup{\AA}$).\\

\newpage
\vskip 0.6 truecm
\par\noindent
%\begin{center}{\bf Reference} \end{center}
\vskip 0.3 truecm

\renewcommand{\baselinestretch}{0.2}

\end{document}